# Compact laser system for a laser-cooled ytterbium ion microwave frequency standard

S. Mulholland[1,2], H. A. Klein[1], G.P. Barwood[1,a], S. Donnellan[1], P.B.R. Nisbet-Jones[1], G. Huang[1], G. Walsh[1], P.E.G. Baird[2] and P. Gill[1,2]

[1] National Physical Laboratory, Teddington, TW11 0LW, UK
[2] Clarendon Laboratory, University of Oxford, Oxford, UK
[a] Corresponding author email: geoffrey.barwood@npl.co.uk

Abstract

The development of a transportable microwave frequency standard based on the ground-state transition of $^{171}$Yb$^+$ at ~12.6 GHz requires a compact laser system for cooling the ions, clearing out of long-lived states and also for photoionisation. In this paper, we describe the development of a suitable compact laser system based on a 6U height rack-mounted arrangement with overall dimensions 260 × 194 × 335 mm. Laser outputs at 369 nm (for cooling), 399 nm (photoionisation), 935 nm (repumping) and 760 nm (state clearout) are combined in a fiber arrangement for delivery to our linear ion trap and we demonstrate this system by cooling of $^{171}$Yb$^+$ ions. Additionally, we demonstrate that the lasers at 935 nm and 760 nm are close in frequency to water vapor and oxygen absorption lines respectively; specifically, at 760 nm, we show that one $^{171}$Yb$^+$ transition is within the pressure broadened profile of an oxygen line. These molecular transitions form convenient wavelength references for the stabilization of lasers for a $^{171}$Yb$^+$ frequency standard.

Introduction

Compact and transportable microwave frequency standards are currently based on beam-tubes [1], vapor cells [2, 3] and buffer-gas cooled trapped ions [4-6]. Large laboratory-based microwave frequency standards based on cesium fountains achieve the highest performance, with typical evaluated uncertainties in the low parts in $10^{16}$ region [7-9] using laser-cooled atoms. Our recently-developed microwave frequency standard is referenced to a 12.6-GHz transition observed in cold $^{171}$Yb$^+$ ions confined in a linear trap. This system [10] is compact and transportable and intermediate in terms of complexity and frequency stability between commercial and laboratory devices. It is designed to operate in a research laboratory environment and future development will focus on increased operational robustness.

As part of our compact rack-mounted frequency standard, we have developed a laser system that sits within the complete unit. In addition to the application to our microwave frequency standard, these lasers could additionally be used as part of a single $^{171}$Yb$^+$ ion transportable optical frequency standard with additional lasers driving either the quadrupole transition at 436 nm or the octupole transition at 467 nm [11, 12]. For the microwave frequency standard, four lasers are required to drive the transitions that are shown in the term scheme in figure 1. The principal cooling transition in $^{171}$Yb$^+$ is at 369.5 nm; fluorescence at this wavelength is also used for state detection. This wavelength, together with a source at 399 nm tuned to a transition in neutral $^{171}$Yb, is also used for photoionisation [13] to create the ions from an atomic beam produced using an oven. Finally, two lasers at 935 nm and 760 nm are used to drive ions out of long-lived $^2D_{3/2}$ and $^2F_{7/2}$ states respectively. We opted to clear-out from the F state via the transition at 760 nm [14] because it is more efficient than an alternative line at 638 nm [15]; 760 nm lasers are also more widely available and more compact than at 638 nm. A photograph of the lasers mounted within their 6U height sub-rack (260 × 194 × 335 mm) is shown in figure 2.

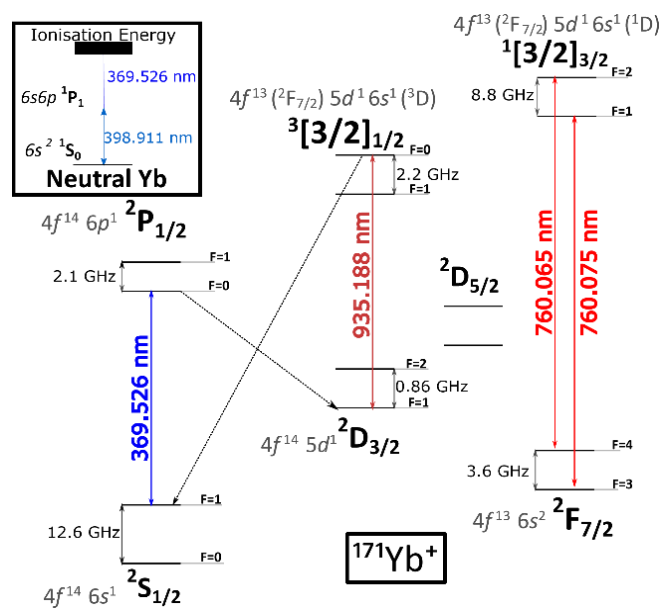

**Figure 1:** $^{171}$Yb$^+$ partial term scheme showing the transitions and the laser wavelengths required for cooling, together with the lasers required for clear-out from the $^2D_{3/2}$ and $^2F_{7/2}$ long-lived states; the microwave clock transition at ~12.6 GHz is also shown. In the figure, laser-driven transitions are shown as colored arrows; black arrows represent probable decay channels. The inset shows the wavelengths in neutral ytterbium used for photoionisation.

Of the four lasers required, the two ultra-violet lasers at 369.5 nm and 399 nm were the most challenging to design in a compact format. To provide the cooling light at 369.5 nm, we used a commercial GaN-based extended cavity diode laser (ECDL). This could provide up to 4 mW output and was the largest and heaviest component of our laser system. Extended cavity lasers, even when well-designed, have long-term mechanical instabilities that can give rise to mode hops. This makes this solution inappropriate for long-term use in an arrangement that is ultimately expected to operate with minimal user intervention.

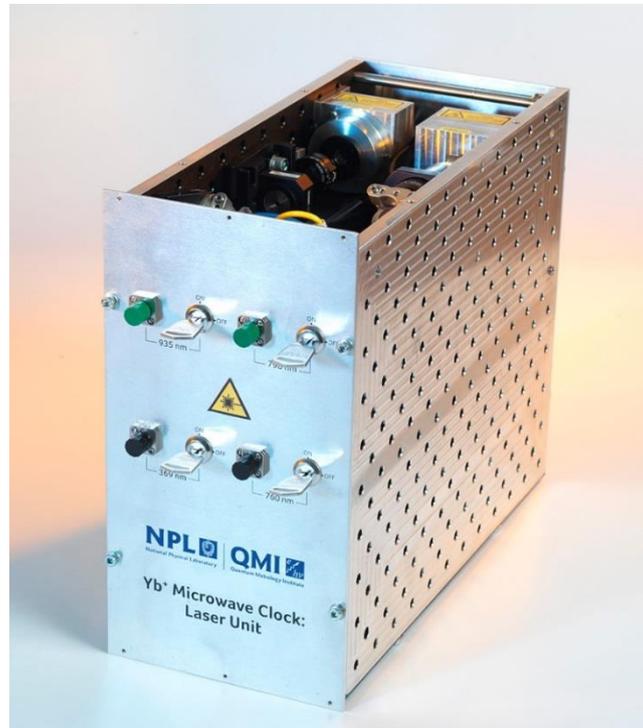

**Figure 2:** Photograph of the complete laser system, packaged in a 6U height sub-rack (260 × 194 × 335 mm).

Our long-term aim is therefore to replace the 369-nm laser with a smaller and more robust unit and, for this reason, we evaluated frequency doubling of a bespoke commercially-available 739-nm diode laser [16] providing up to 60 mW output in free space. Using an AdvR waveguide frequency doubler with a pig-tailed input and free-space output, ~24 μW was produced at 369.5 nm from 15 mW in the polarization-maintaining input fiber. With further optimization of the fiber input coupling and minimization of the transmission losses between the laser and doubler, we expect that up to 100 μW could be produced for 30 mW in fiber at 739 nm. However, for our current trap, we require a minimum of 100 μW for photoionisation [13] and efficient cooling and so our doubler efficiency is currently marginal compared with requirements. However, rapid progress is being made in the commercial availability of doublers and lasers at 739 nm and we foresee this route as being viable for $^{171}$Yb$^+$ cooling in the near future.

Light at 399 nm for photoionisation of neutral ytterbium atoms was generated by frequency doubling of a commercial diode laser at 798 nm [16] that provided up to 80 mW in free space. The laser had an integral thermo-electric cooler in a TO-39 package that included a collimator and also a miniature volume holographic grating (VHG) for frequency control. The output was coupled via a free-space Faraday isolator into polarization maintaining fiber. Frequency doubling was provided by a pig-tailed waveguide doubler and ~170 μW output was obtained in the output fiber at 399 nm for ~50 mW in the fiber input at 798 nm. This doubler was supplied to us by AdvR [17] and is similar to that recently demonstrated by a group at the National Metrology Institute of Japan in collaboration with the NTT Electronics Corporation [18].

The lasers at 798 nm, 935 nm and 760 nm were all driven using commercially available prototype boards incorporated into our frequency standard. The 369.5-nm laser required a higher compliance voltage and therefore, for this device, we used an in-house laser driver design. NPL-designed electronics were also used for the temperature controllers required for the lasers and the frequency doubler. These digital controllers were interfaced to a personal computer (PC) for control of the temperature control point and individual gain settings for the proportional-integral-derivative (PID) parts of the servo system. All the electronics, including both in-house NPL designed and commercially-sourced circuits were housed in a 6U height unit [10]. The top-level schematic of the arrangement inside this unit is shown in figure 3.

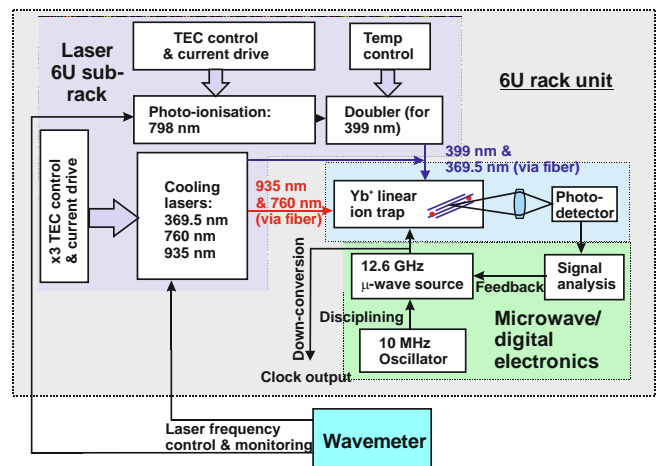

**Figure 3:** Schematic of the complete microwave clock assembly [10], including the lasers, microwave and digital electronics and ion trap. All the components except for the wavemeter are housed within a 6U height rack.

Fiber input to the trap and cold $^{171}$Yb$^+$ ion spectroscopy

The detailed optical arrangement for combining the outputs from all four lasers and focusing them into the trap is shown in figure 4. Low-power pick-offs from the infrared lasers at 760 nm and 935 nm were sent to a wavemeter via a fiber switcher and the higher-power outputs were combined in a common polarization maintaining input fiber to a free space collimator fixed to a trap window. The two ultra-violet sources at 369.5 nm and 399 nm were combined in free space and then directed via a second polarization maintaining fiber to a collimator attached to a different trap window. The 369.5 nm frequency was monitored directly by the wavemeter using a low-power pick-off; the 399 nm frequency was monitored via a pick-off at 798 nm. At both 399 nm and 369.5 nm, mechanical shutters were used; switching for the cooling laser was required as part of the measurement cycle for locking to the clock transition [10]. The 399-nm shutter required for photoionisation was only opened when the trap required loading. Finally, an optical attenuator (10% transmission) was provided on a relay arrangement that allowed the cooling beam to be sent to the trap either in "full power" or "low power" mode under computer control. The high cooling beam power is used for ion loading, cooling and state preparation and the lower power for state detection of the ions.

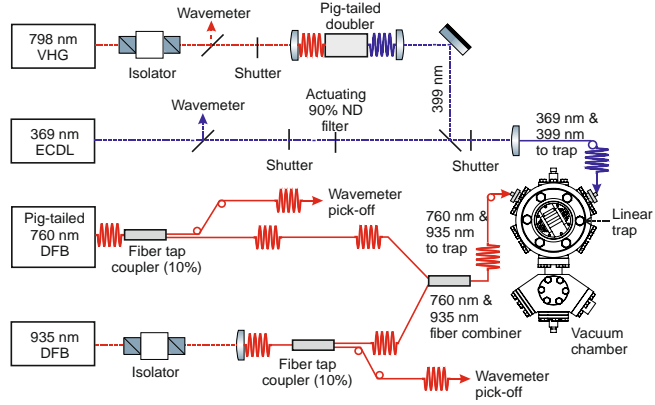

**Figure 4:** Overall laser schematic showing how the ultra-violet and infrared beams are combined in fiber in our laser system and then coupled into the trap. Both the 760 nm and 935 nm sources are distributed feedback (DFB) lasers; 369 nm is produced by an extended cavity diode laser (ECDL) and the 798 nm [16] comprises a diode laser with optical feedback from a miniature volume holographic grating (VHG). Free-space beam paths are indicated by dashed lines; solid lines represent fiber links. The 369 nm and 760 nm laser packages come with integral optical isolators and therefore no further external isolators are required. The cooling beam is directed along the axis of the linear trap; the infrared clear-out light is at 120° to this in the trap.

The arrangement shown in figure 4 provided 300 µW of 369.5 nm light at the trap, focused to a ~200 µm beam waist (1/e intensity radius); around 50 µW at 399 nm focused to ~200 µm was also available when loading. The infrared lasers were not switched but remained on at the trap continuously; we estimated that the presence of these beams at 760 nm and 935 nm caused a negligible Stark shift on the 12.6-GHz microwave transition. The 760-nm laser provides ~0.5 mW focused to ~300 µm; a much higher power (3 mW) is used at 935 nm, focused to a similar beam waist.

For optimum performance, frequency stabilization of each of our lasers was required and this was achieved using a HighFinesse WS10 wavemeter housed separately from the main unit. A Zeeman stabilized HeNe laser [19] provided a local frequency reference for this wavemeter and frequency control was achieved using algorithms similar to [20]. The required level of frequency stability is different for each laser; for the 369.5 nm cooling transition the stability needs to be significantly better than the 19.6-MHz natural linewidth of the cooling transition [21, 22]. Prior to loading, the laser is detuned by typically 100 MHz to the low-frequency side of the transition. This initial cooling laser frequency control is via the wavemeter and, once a fluorescence signal is observed, this fluorescence can also be used for long-term laser frequency monitoring. Less stringent frequency control is required for the other sources since the 935-nm and 760-nm transitions are power broadened and the 399-nm photoionisation laser requires control only within the Doppler-broadened profile observed when the atomic beam exits the oven.

The $^{171}$Yb$^+$ ions were confined in a linear trap comprising four 2-mm rods with a center-to-center spacing of 7.3 mm and five plate electrodes; together, these provide both radial and axial confinement and the complete trap geometry is described in [10]. In figure 5, we present a cooling profile observed with a cloud of $^{171}$Yb$^+$ ions using our compact cooling laser system. Without the 935 nm laser, the ions are driven into the D$_{3/2}$ levels (figure 1) with a probability of ~1/200 [23] and no longer fluoresce. The lower trace of figure 5 demonstrates this by plotting the ion fluorescence as a function of 935 nm laser de-tuning. The fitted ~115 MHz half width at half maximum (HWHM) linewidth demonstrates the power broadening of this transition which relaxes the laser frequency stability requirement.

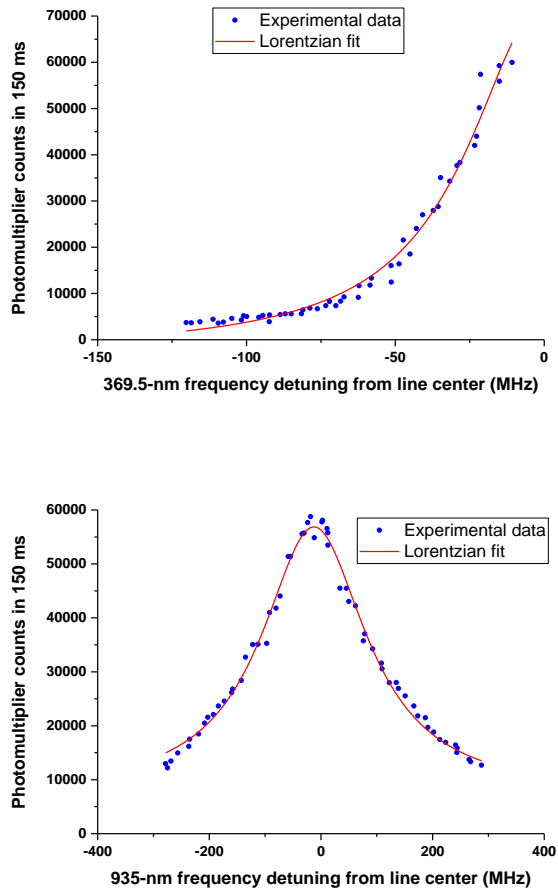

**Figure 5:** Observation of the $^{171}$Yb$^+$ cooling profile (upper trace) and ion fluorescence observed from a cloud of ions as a function of 935 nm laser detuning (lower trace) using our compact laser system and linear trap. Both fluorescence datasets have been fitted to Lorentzian profiles although, in the upper plot, the fit was constrained to have a fluorescence peak at zero frequency detuning. The fitted HWHM of the cooling profile is ~30 MHz; for the profile obtained by scanning the 935 nm laser, the fitted HWHM was ~115 MHz.

Near infrared laser spectroscopy of water and oxygen

Of the four lasers required, the most straightforward to source are those at 935 nm and 760 nm. These are distributed feedback (DFB) sources that are commercially packaged with integral thermo-electric coolers. They are marketed for use in water vapor and oxygen detection respectively and, in each case, the relevant laser source can temperature tune both to the required $^{171}$Yb$^+$ transitions and the molecular transitions. This means that both 760 nm and 935 nm lasers can also be used as a frequency reference independent of $^{171}$Yb$^+$ as a reference for a wavemeter measuring the 399 nm or 369.5 nm laser light. In a future development, this is a more compact solution than our Zeeman stabilized HeNe laser.

We have made measurements of the frequencies of the $^2F_{7/2}$ (F = 4) → $^1[3/2]_{3/2}$ (F = 2) and $^2F_{7/2}$ (F = 3) → $^1[3/2]_{3/2}$ (F = 1) transitions in $^{171}$Yb$^+$ (figure 1). These were made using a single $^{171}$Yb$^+$ ion confined in a Paul trap that is configured as an optical frequency standard. This allowed us to drive the ion into one of the $^2F_{7/2}$ states using light at 467 nm [11] which stopped the ion fluorescing and then tuning the 760-nm laser until fluorescence recovery was observed. Measurements were made using a High Finesse WS10 wavemeter, calibrated with a laser at 698 nm used in the NPL strontium lattice clock [24, 25]. These measurements yielded wavelengths of 760.065(1) nm and 760.075(1) nm for the above two transitions respectively, where the measurement standard uncertainties are indicated in parenthesis. Specifically, these measurements show that there is overlap between the pressure-broadened transition in oxygen at 13156.623 cm$^{-1}$ ($\equiv$ 760.073 nm) tabulated in the HITRAN16 [26] molecular absorption database and the $^2F_{7/2}$ (F = 3) $\rightarrow$ $^1[3/2]_{3/2}$ (F = 1) transition in $^{171}$Yb$^+$ at 760.075 nm.

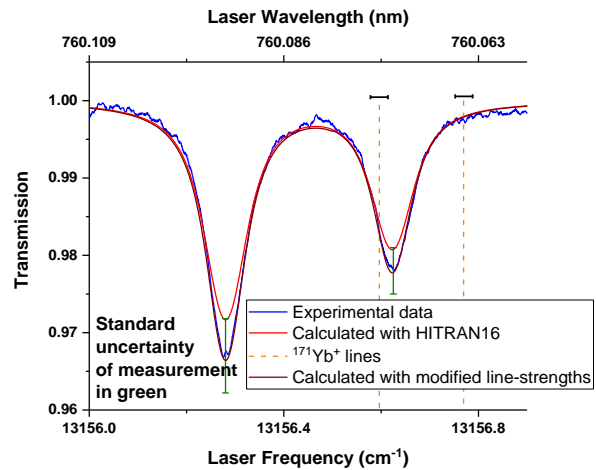

**Figure 6:** Oxygen data and the transition frequencies of $^{171}$Yb$^+$ at 760 nm. The $^{171}$Yb$^+$ measurement uncertainties are indicated in black in the above plot. The results have been corrected for the detector non-linearity and the green vertical error bar shows our estimated one standard deviation uncertainty estimate for oxygen maximum linear absorption. The linear absorption profile is also shown, calculated using the HITRAN16 tabulated line-strengths. However, optimum fitting required using line-strengths of 1.13 and 1.10 times the tabulated values for the higher and lower frequency lines respectively; optimum fitting also required a pressure broadening coefficient for both lines of 0.971 of the tabulated value (table 1).

Since the rate at which we drive into these F-states is low, we use one 760-nm laser that is driven with a typical 10-s period and a square-wave frequency amplitude of 5.2 GHz, which is the combined ground and upper state splitting (figure 1). Figure 6 shows the calculated linear absorption of oxygen over a ~1.6 m air path at standard temperature and pressure; the calculated absorption using a Voigt profile and we use HITRAN16 values for the air-pressure broadening coefficients.

Although not critical for our use as a wavelength reference, we have made measurements of the linear absorption of oxygen at 760 nm and our results are also shown in figure 6. Measurement of linear absorption at the few percent level requires careful characterization of the photodiode used. For example, detection of the peak absorption that is typically only 3%, and measured to an uncertainty of 10% of this value, means that the photodiode and pre-amplifier need to be linear to better than 0.3%. Our detector was linear only at the ~1% level but, by making measurements at different optical power levels and characterizing the detector response as a function of power, we were able to correct for this non-linearity. A comparison of the measured values of line-strength with those from HITRAN16 is given in table 1. This table also includes values for the pressure-broadened molecular transition linewidths. Determining line center to better than ~5% of this HWHM (i.e. ~60 MHz) is expected to provide the required laser frequency accuracy. These transitions could provide a wavemeter frequency reference and a full characterization of the wavemeter is expected to allow us to realize a similar level of uncertainty for all four lasers prior to ion loading. This uncertainty is less than the power-broadened 935-nm linewidth of 115 MHz (figure 5) and, also less than the 369.5-nm laser de-tuning of ~100 MHz applied during ion loading.

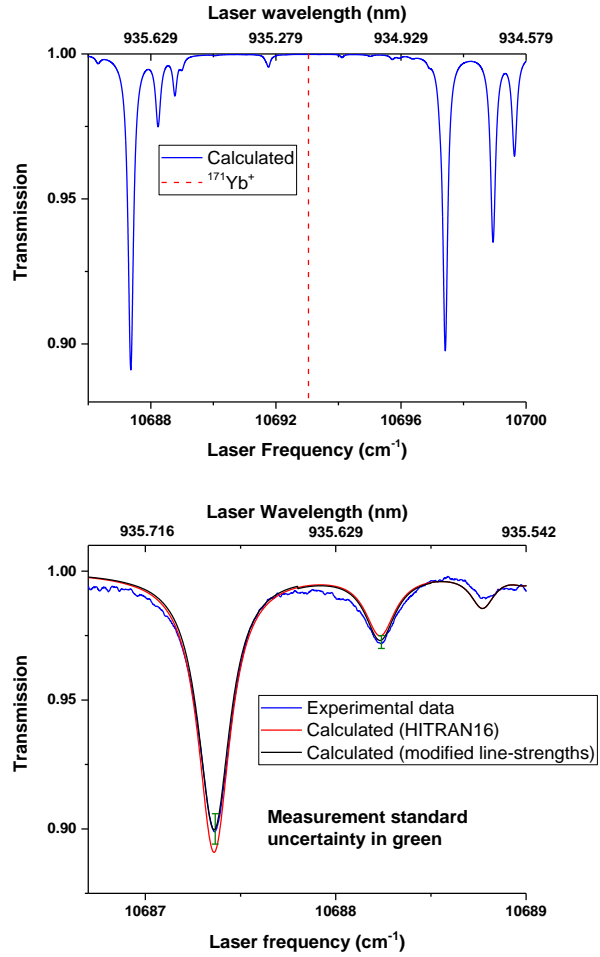

**Figure 7:** Calculated and measured water vapor absorption lines over a 1.6-m air path and the $^{171}$Yb$^+$ transition (upper plot) for a temperature of 20.1°C and relative humidity of 52%, which were the environmental conditions prevailing at the time of our absorption measurements (lower plot). Lasers purchased for use at the $^{171}$Yb$^+$ transition frequency can also be expected to tune over either the higher or lower frequency water absorption lines, depending on the center wavelength of the laser. Our laser was able to access the lower frequency set of lines. The results from the measurement have been corrected for the detector non-linearity and the green vertical error bar shows our estimated one standard deviation uncertainty. We also show the linear absorption profile calculated using the HITRAN16 tabulated line-strengths. In this case, optimum fitting required correcting the tabulated line-strengths for the two strongest absorptions by factors of 0.92 (for the lowest-frequency line) and 1.07. The observed linewidth does not differ within experimental uncertainty from that predicted by HITRAN16.

|  | Frequency (cm$^{-1}$) | HITRAN16 line-strength cm$^{-1}$/(mol cm$^2$) | Measured line-strength cm$^{-1}$/(mol cm$^2$) | Pressure broadening (HWHM; cm$^{-1}$/atm) |
|---|---|---|---|---|
| O$_2$ | 13156.280183 | 4.799 x 10$^{-24}$ | 5.4(0.9) x 10$^{-24}$ | 0.0433 (HITRAN16 value 0.04460) |
| O$_2$ | 13156.623154 | 3.123 x 10$^{-24}$ | 3.3(0.6) x 10$^{-24}$ | 0.0433 (HITRAN16 value 0.04460) |
| H$_2$O | 10687.361176 | 6.496 x 10$^{-22}$ | 6.0(0.4) x 10$^{-22}$ | 0.09530 (as listed in HITRAN16) |
| H$_2$O | 10688.233180 | 1.284 x 10$^{-22}$ | 1.4(0.1) x 10$^{-22}$ | 0.09080 (as listed in HITRAN16) |

Table 1: Table of HITRAN16 line-strengths for the four strongest observed molecular lines; two in oxygen at 760 nm and two in water vapor at 935 nm, together with our measured values.

From the upper level of the 369.5 nm cooling transition, a $^{171}$Yb$^+$ ion can decay to a long-lived [27] $^2$D$_{3/2}$ state. In order to return the ion to the cooling cycle, a further laser is required that is tuned to the $^2$D$_{3/2}$ (F = 1) → $^3$[3/2]$_{1/2}$ (F = 0) transition; the required vacuum wavelength is 935.18768(19) nm [28]. The overlap with nearby water vapor absorptions at 935 nm is not so close in frequency as at 760 nm, but the same laser purchased to tune to the $^{171}$Yb$^+$ 935-nm transition can be expected to temperature tune to strong water vapor absorption lines either higher or lower in frequency, depending on the individual laser tuning properties (figure 7; upper plot). Our laser could tune to transitions lower in frequency than the $^{171}$Yb$^+$ transition and, in figure 7, we show calculated water vapor absorption profiles over a 1.6-m path length at 52(2)% relative humidity and 20.1(0.1)°C. These parameters have been converted to a partial pressure of water vapor using the Buck equation [29, 30] which gives a water vapor content in air of 1.21(5)% by volume; measurement standard uncertainties are given in parentheses. This uncertainty in the determination of the partial pressure of water vapor is therefore ~4%. As already discussed in the previous paragraph, we need to take account of detector non-linearity when measuring linear absorptions that are at the few percent level. By measuring the concentration at different power levels on our detector and characterizing the output voltage as a function of laser power, we were also able to correct for this non-linearity resulting in a further uncertainty of 4.4% to the total linear absorption. Combining the uncertainties in quadrature from detector non-linearity and water vapor partial pressure determination gives a total uncertainty of 6% in the value of our measured peak absorption and this error bar is indicated in figure 7.

For both the oxygen and water vapor linear absorption data, we fitted the results using an assumed Voigt profile that was calculated in terms of the real part of the complex Fadeeva function $w(z)$ where $z$ is the complex parameter defined by [31, 32]:

$$z = \frac{\Delta \nu + j\gamma}{\nu_D} \sqrt{\ln 2} \tag{1}$$

Here, $\gamma$ is the pressure broadened HWHM (i.e., the Lorentzian component of the Voigt profile) and $\nu_D$ is the Doppler (Gaussian component) HWHM. The Doppler component is calculated to be 0.0142 cm$^{-1}$ ($\equiv$ 426 MHz) for oxygen and 0.0154 cm$^{-1}$ ($\equiv$ 461 MHz) for water at 293 K.

Summary

We have described a compact laser system designed and built for a transportable turn-key microwave frequency standard at ~12.6 GHz, using cooled ions confined in a linear trap. This laser system could additionally be used for cooling and photoionisation in a $^{171}$Yb$^+$ ion compact optical frequency standard using either the quadrupole transition at 436 nm or the octupole transition at 467 nm. We also explore an alternative solution using commercially available components for generating 369.5 nm by frequency doubling rather than a direct diode output. Frequency doubling to 369.5 nm has the dual benefits of being more compact than our extended cavity laser system and, by moving all our laser sources to the infrared, increases options for simpler wavelength stabilization. In our current system, this is provided by a wavemeter external to the main device. However, in a future system, we need to source a more compact wavemeter that could be mounted within the laser unit. We have evaluated a newly available commercial compact wavemeter [33], operational from 700 nm to 1000 nm that neatly overlaps our requirements to measure 760 nm, 935 nm and the two ultra-violet wavelengths when generated by frequency doubling. The reproducibility of this new commercial device has been found to meet our requirement to provide laser frequency measurements independent of an ion fluorescence signal. Finally, we demonstrate that both the 760 nm and 935 nm lasers have convenient nearby molecular absorption frequency references that could be used to calibrate the wavemeter performance.

Acknowledgements

The authors would like to gratefully acknowledge UK Defence Science and Technology Laboratory (Dstl) and also Innovate UK for funding the work reported here. We would also like to thank our NPL colleagues Pravin Patel for advice on the electronics, Steven King for early work on the vacuum system and optics design and colleagues from the NPL ytterbium-ion and strontium lattice optical clock projects for their help in making the wavelength measurements at 760 nm. Finally, we thank Anne Curtis for her careful reading of the manuscript.